\documentclass[journal]{IEEEtran}

\usepackage[dvipsnames]{xcolor}
\usepackage[acronym]{glossaries-extra}
\usepackage[colorlinks, citecolor=blue, linkcolor=blue]{hyperref}
\usepackage[textwidth=1.6cm]{todonotes}
\usepackage{comment}
\usepackage{nameref}
\usepackage{subcaption}
\usepackage{multicol}
\usepackage{multirow}
\usepackage{rotating}
\usepackage{stfloats}
\usepackage[noadjust]{cite}
\usepackage{float}
\usepackage{amsfonts}
\usepackage{upgreek}
\usepackage{siunitx}
\usepackage{mathtools}
\usepackage{scalerel}
\usepackage{braket}
\usepackage{soul}
\usepackage{tcolorbox}
\usepackage{lipsum}

\newcommand{\Gr}{g_{\mathrm{R}}}
\newcommand{\GR}{G_{\mathrm{R}}}
\newcommand{\Cr}{c_{\mathrm{R}}}
\newcommand{\Fr}{\vec{f}_{\mathrm{R}}}
\newcommand{\Frm}[1]{f_{\mathrm{R,#1}}}

\newcommand{\eff}{\mathrm{eff}}

\newcommand{\Aeff}[1]{A_{\mathrm{eff},#1}}
\newcommand{\Aeffm}[1]{\hat{A}_{\mathrm{eff},#1}}
\newcommand{\LeffMat}{\mathbf{L}_{\mathrm{eff}}}
\newcommand{\AeffMat}{\mathbf{A}_{\mathrm{eff}}^{\mathrm{inv}}}
\newcommand{\R}{\mathrm{R}}
\newcommand{\rf}{\mathrm{ref}}

\newcommand{\der}{\mathrm{d}}
\newcommand{\ad}{\mathrm{ad}}
\newcommand{\diag}{\mathrm{dg}}

\newcommand{\OSNR}{\mathrm{OSNR}}

\newcommand{\fmax}{f_{\mathrm{Max}}}
\newcommand{\fmin}{f_{\mathrm{Min}}}

\newcommand{\Th}{\mathrm{th}}

\newcommand{\Tx}{\mathrm{Tx}}
\newcommand{\Rx}{\mathrm{Rx}}
\newcommand{\Total}{\mathrm{T}}

\newcommand{\dB}{\mathrm{dB}}

\let\bigopsize\bigoplus
\def\bigoplus{{\scalerel*{\boldsymbol\oplus}{\bigopsize}}}

\newcommand{\appref}[1]{\hyperref[#1]{Appendix~\ref*{#1}}}

\makeatletter
\newcommand{\subalign}[1]{%
  \vcenter{%
    \Let@ \restore@math@cr \default@tag
    \baselineskip\fontdimen10 \scriptfont\tw@
    \advance\baselineskip\fontdimen12 \scriptfont\tw@
    \lineskip\thr@@\fontdimen8 \scriptfont\thr@@
    \lineskiplimit\lineskip
    \ialign{\hfil$\m@th\scriptstyle##$&$\m@th\scriptstyle{}##$\hfil\crcr
      #1\crcr
    }%
  }%
}
\makeatother

\setabbreviationstyle[acronym]{long-short}

\newacronym{snr}{SNR}{signal-to-noise ratio}
\newacronym{osnr}{OSNR}{optical signal-to-noise ratio}
\newacronym{sinr}{SINR}{signal-to-interference-plus-noise ratio}

\newacronym{smf}{SMF}{single-mode fiber}
\newacronym{sdm}{SDM}{space-division multiplexing}
\newacronym{ssmf}{SSMF}{standard single-mode fiber}
\newacronym{mmf}{MMF}{multi-mode fiber}
\newacronym{si-mmf}{SI-MMF}{step-index multi-mode fiber}
\newacronym{gi-mmf}{GI-MMF}{graded-index multi-mode fiber}
\newacronym{mcf}{MCF}{multi-core fiber}
\newacronym{uc-mcf}{UC-MCF}{uncoupled multi-core fiber}
\newacronym{cc-mcf}{CC-MCF}{coupled multi-core fiber}

\newacronym{dra}{DRA}{distributed Raman amplification}
\newacronym{mdg}{MDG}{mode-dependent gain}
\newacronym{mdl}{MDL}{mode-dependent loss}
\newacronym{csi}{CSI}{channel state information}
\newacronym{cdf}{CDF}{cumulative distribution function}
\newacronym{pdf}{PDF}{probability density function}
\newacronym{gue}{GUE}{Gaussian unitary ensemble}
\newacronym{wdm}{WDM}{wavelength-division multiplexed}
\newacronym{mmse}{MMSE}{minimum mean squared error}
\newacronym{mimo}{MIMO}{multiple-input multiple-output}
\newacronym{msle}{MSLE}{mean squared logarithmic error}
\newacronym{awgn}{AWGN}{additive white Gaussian noise}
\newacronym{rmse}{RMSE}{root-mean-squared}
\newacronym{nli}{NLI}{non-linear interference}
\newacronym{gn}{GN}{Gaussian noise}
\newacronym{dge}{DGE}{dynamic gain equalizer}
\newacronym{cw}{CW}{continuous wave}

\newacronym{lp}{LP}{linearly-polarized}
\newacronym{qkd}{QKD}{quantum-key distribution}
\newacronym{psd}{PSD}{power spectral density}
\newacronym{ase}{ASE}{amplified spontaneous emission}
\newacronym{srs}{SRS}{stimulated Raman scattering}
\newacronym{isrs}{ISRS}{inter-channel stimulated Raman scattering}
\newacronym{sprs}{SpRS}{spontaneous Raman scattering}
\newacronym{fwm}{FWM}{four-wave-mixing}
\newacronym{imxt}{IMXT}{inter-mode cross-talk}

\begin{document}
\title{Raman amplification and ISRS in SDM links: Analytical evaluation and closed-form models for optical transmission}
\author{Lucas~Alves~Zischler,~\IEEEmembership{Student~Member,~IEEE,}%
        ~Antonio~Mecozzi,~\IEEEmembership{Fellow,~IEEE,}
        ~and~Cristian~Antonelli,~\IEEEmembership{Senior~Member,~IEEE}
\thanks{Manuscript received XXX xx, XXXX; revised XXXXX xx, XXXX; accepted XXXX XX, XXXX. Funded by the European Union (Grant Agreement 101120422). Views and opinions expressed are, however, those of the author(s) only and do not necessarily reflect those of the European Union or REA. Neither the European Union nor the granting authority can be held responsible for them. (Corresponding Author: Lucas~Alves~Zischler)}%
\thanks{Lucas~Alves~Zischler, Antonio~Mecozzi, and Cristian~Antonelli are with the Department of Physical and Chemical Sciences, University of L’Aquila, 67100 L’Aquila, Italy: (\mbox{e-mail: lucas.zischler@univaq.it}).}}%

\markboth{}%
{Raman amplification and ISRS in SDM links: Analytical evaluation and closed-form models for optical transmission}

\maketitle

\begin{abstract}
  In optical communications, the Raman effect is exploited for its lasing properties in \gls*{dra} and leads to spectral distortions through \gls*{isrs}. In single-mode fibers, these effects are well understood and modeled, but equivalent closed-form expressions for arbitrarily coupled \gls*{sdm} links are lacking. In this work, we expand upon previous literature by providing closed-form expressions modelling \gls*{dra} and \gls*{isrs} in common \gls*{sdm} fiber designs that support arbitrarily-coupled degenerate mode-groups, incorporate mode coupling, and accounting for inter-modal non-linear effects, showing excellent agreement with simulations. The derived formulas are then applied to representative scenarios, illustrating how distinct pump and fiber configurations influence gain and \gls*{mdg}. Finally, we describe suggested routines for experimentally estimating the Raman response profiles of \gls*{sdm} fibers.
\end{abstract}

\glsresetall

\begin{IEEEkeywords}
  Non-linear effects, Propagation model, Space-division multiplexing (SDM), Distributed Raman amplification (DRA), Inter-channel stimulated Raman scattering (ISRS)
\end{IEEEkeywords}

\glsresetall

\section{Introduction}

\IEEEPARstart{T}{he} last decade of optical communication research has seen a significant increase in interest toward novel fiber designs beyond the \gls*{ssmf}. In particular, \gls*{sdm} has been in the spotlight due to its ability to achieve high throughput within standard cladding dimensions~\cite{winzer2012optical}. It has also attracted attention in other optical fields such as quantum key distribution~\cite{ding2017high} and sensing~\cite{li2015few}, becoming a commonly discussed subject. Likewise, the Raman effect is present in all the previously mentioned areas. In optical transmission, \gls*{srs} enables \gls*{dra}~\cite{bromage2004raman}, and \gls*{srs} interactions between channels induce a tilting of the power spectrum due to the transfer of power from higher-frequency channels to lower-frequency ones via \gls*{isrs}~\cite{zirngibl1998analytical}. In quantum key distribution and sensing applications, \gls*{sprs} is often discussed either as an adverse effect that corrupts weak quantum signals~\cite{peters2009dense} or as a mechanism for probing the optical fiber, such as in temperature measurements~\cite{dakin1985distributed}.

As seen throughout the literature, both phenomena are intrinsic to the field of optics, with applications in many areas. Nevertheless, work addressing applications combining both effects remains sparse. With the rising interest in \gls*{sdm} transmission, efforts have been made to extend \gls*{dra} to this new domain. Notably, in~\cite{ryf2012analysis,antonelli2013raman} the authors provide semi-analytical frameworks describing the power evolution of pumps and signals in fibers supporting multiple degenerate mode-groups. In~\cite{ryf2011mode,rademacher2022investigation} the authors experimentally investigate \gls*{dra} in \glspl*{mmf}. In~\cite{sciullo2025characterization} the Raman gain is characterized in \glspl*{mcf}. In~\cite{schneck2025experimental,zischler2025evaluation} \gls*{isrs} is studied both experimentally and analytically in \glspl*{mmf}. In~\cite{zischler2025spontaneous} we investigate \gls*{sprs} in \gls*{sdm} fibers. Nevertheless, no work has since provided simplified closed-form expressions which model Raman effects in arbitrarily-coupled \gls*{sdm} systems. In addition, a clear formalism is needed for scenarios involving combined \gls*{srs} and crosstalk interactions, including the derivation of quantitative metrics to compare different Raman pump schemes.

In this work, we analytically evaluate the behavior of Raman effects in \gls*{sdm} systems, expanding the analytical contributions of our previous works~\cite{sciullo2025characterization,zischler2025evaluation,zischler2025spontaneous}, which neglected coupling between mode-groups. We encompass Raman amplification, its associated \gls*{ase} noise, and \gls*{isrs}. We discuss how the Raman effect behaves in \glspl*{mcf} and \glspl*{mmf}, and how core spacing and distinct doping profiles influence the resulting Raman gain. We present simple closed-form expressions to model the evolution of signal powers. Furthermore, we establish a formalism for the on–off Raman gain matrix and \gls*{mdg} and derive its respective analytical expressions. Finally, we provide suggested routines and expressions for experimentally characterizing the fiber Raman response in coupled regimes.

The work is structured as follows. Section~\ref{sec:ii} provides an overview of the Raman effect and its behavior in fiber types supporting multiple orthogonal spatial channels. Section~\ref{sec:iii} presents the main contributions of this work, including analytical models for \gls*{dra} and \gls*{isrs}, as well as semi-analytical models dictating Raman \gls*{ase} noise in arbitrarily-coupled \gls*{sdm} fibers. Section~\ref{sec:iv} applies the derived models to representative scenarios, comparing distinct pump schemes and fiber types. Section~\ref{sec:v} introduces frameworks for characterizing Raman efficiency in \gls*{sdm} fibers. Finally, Section~\ref{sec:vi} concludes the paper.

\section{The Raman effect in SDM fibers}
\label{sec:ii}

The Raman effect was first observed by $\text{Raman}$ and $\text{Krishnan}$ in~\cite{raman1928new}, who measured a scattered response from the medium at lower energy levels than the incident photons, indicating an energy loss due to vibrational absorption within the material. In addition to \gls*{sprs}, the Raman effect enables lasing within optical fibers, amplifying lower-frequency signals through \gls*{srs}.

In optical transmission, the Raman response of silica can be considered instantaneous on the time scale of typical signal pulses~\cite{lin2006raman}. Consequently, \gls*{srs} can be modeled as an instantaneous transfer of power from higher to lower frequencies. This effect is observed in two distinct ways in optical transmission: Raman amplification, in which strong unmodulated pump signals are deliberately placed at higher frequencies to amplify desired channels, and \gls*{isrs}, in which data channels at higher frequencies transfer power to those at lower frequencies, producing a characteristic tilt in the received power spectrum.

Both effects are well understood and accurately modeled in \gls*{smf} systems~\cite{bromage2004raman, zirngibl1998analytical}. However, the distribution of the electric fields among the multiple orthogonal mode profiles of \gls*{sdm} fibers, together with mode-coupling effects, results in intricate Raman interactions, that complicate the derivation of simple analytical models.

\subsection{SDM fiber types}

\begin{figure}[!t]
    \centering
    \includegraphics{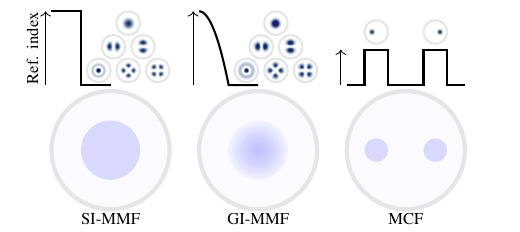}
    \caption{Refractive index profiles, lateral field power distributions, and cross-sections of a \gls*{si-mmf}, a \gls*{gi-mmf}, and a 2-core \gls*{mcf}.}
    \label{fig:FiberTypes}
\end{figure}

The additional orthogonal spatial paths provided by \gls*{sdm} fibers can take several forms. Currently, two \gls*{sdm} fiber types are highlighted in the literature: \glspl*{mcf} and \glspl*{mmf}, with representative designs shown in Fig.~\ref{fig:FiberTypes}. In \glspl*{mcf}, the field power is concentrated within the individual cores, but a portion of the modes extends into the cladding and may reach adjacent cores, resulting in crosstalk. Depending on the design, an \gls*{mcf} may be classified as a \gls*{cc-mcf}, in which the cores experience strong mixing and behave as a single degenerate hyper-polarized state, or as an \gls*{uc-mcf}, in which the fields interact only weakly and the polarization states of each core can be treated as distinct mode-groups.

In \glspl*{mmf}, the mode fields occupy the same core. Nevertheless, the spatial profiles of the modes are mutually orthogonal in the sense that their field overlap integral is zero, in an ideal waveguide. The most common \gls*{mmf} designs are \glspl*{si-mmf}, in which the modes extend over most of the core region, and \glspl*{gi-mmf}, whose parabolic refractive index profile concentrates lower-order modes near the core center, while higher-order modes spread over a wider region. Modes in \glspl*{mmf} can be aggregated into weakly coupled degenerate mode-groups, as those within the same group propagate with similar velocities, increasing crosstalk. Coupling in \glspl*{mmf} is often modeled as arising from perturbations to the waveguide boundaries or refractive index, as discussed in~\cite{marcuse1973coupled}.

As it is often employed in the literature, for the remainder of this work, modes or cores exhibiting strong coupling are grouped into degenerate mode-groups, where individual mode fields cannot be treated independently. Therefore, the behavior of any single mode within a mode-group reflects the behavior of the entire set.

The analysis presented here can be extended to other \gls*{sdm} fiber types, if their lateral mode profiles and coupling characteristics are known.

\subsection{Effective area of non-linear interaction}

Non-linear effects arise from the material response to the local optical field. As the power carried by individual modes is not uniformly distributed across the fiber cross-section, non-linear interactions are concentrated in specific regions. For this reason, discussions of non-linear effects often introduce the concept of ``\textit{effective area}'', which quantifies the spatial extent over which non-linear interactions occur. When considering interactions between distinct mode-field profiles, be it between different cores in \glspl*{mcf} or between spatial modes in \glspl*{mmf}, non-negligible inter-modal non-linear effects can be observed. The strength of these interactions depends on the overlap between the field profiles, as described in~\cite[Eqs.~(6--7)]{poletti2008description}. By analogy with the conventional effective area, we define here a ``\textit{cross-effective area}'' as
\begin{equation}
  \hat{A}_{\text{eff}\hspace{-0.5pt},\hspace{-0.5pt}i\hspace{-0.5pt},\hspace{-0.5pt}h}(f_{1},\hspace{-1pt}f_{2})\hspace{-3pt}=\hspace{-3pt}\frac{\int\hspace{-5pt}\int\hspace{-2pt}||\vec{\mathbf{F}}_{\hspace{-2pt}i}(x,\hspace{-1pt}y,\hspace{-1pt}f_{1})||^{2}\der x\der y\hspace{-2pt}\int\hspace{-5pt}\int\hspace{-2pt}||\vec{\mathbf{F}}_{\hspace{-2pt}h}(x,\hspace{-1pt}y,\hspace{-1pt}f_{2})||^{2}\der x\der y}{\int\hspace{-5pt}\int||\vec{\mathbf{F}}_{i}(x,y,f_{1})\cdot\vec{\mathbf{F}}_{h}(x,y,f_{2})||^{2}\der x\der y}\hspace{-1pt},
\end{equation}
where $\vec{\mathbf{F}}_{i}(x,y,f)$ is the $i^{\Th}$ mode lateral profile at frequency $f$. The value of $\hat{A}_{\text{eff},i,h}$ represents the effective area associated with non-linear interactions between two different modes at possibly different frequencies. The special case $i=h$ reduces to the conventional effective area.

In \glspl*{mcf}, the portion of the mode-field profiles extending into adjacent cores can be often assumed to be negligible, even in \glspl*{cc-mcf}, resulting in negligible inter-modal non-linear interactions. In contrast, as the spatial modes in \glspl*{mmf} occupy the same core, inter-modal non-linear effects are often noticeable, with experimental evidence supporting this observation~\cite{essiambre2013experimental}.

Linear coupling in \glspl*{mmf} is also inversely proportional to the cross-effective area, as described in~\cite[Eq.~(29)]{marcuse1973coupled} and~\cite[Eq.~(4.13)]{okamoto2021fundamentals}, and increase as the differences between the propagation constants of the interacting modes decrease~\cite[Eqs.~(22--24)]{koshiba2011multi},~\cite[Eq.~(13)]{marcuse1972fluctuations}.

Within weakly coupled mode-groups, non-linear interactions can be averaged across the individual modes, leading to the definitions of the \textit{self-} and \textit{cross-effective mode-group areas}
\begin{equation}
  A_{\text{eff},n,m}(f_{1},f_{2})=\frac{D_{n}(f_{1})D_{m}(f_{2})}{\sum_{\substack{i\in\mathbf{M}_{n}(f_{1})\\h\in\mathbf{M}_{m}(f_{2})}}\hat{A}^{-1}_{\text{eff},i,h}(f_{1},f_{2})},
  \label{eq:aeffmg}
\end{equation}
which will be utilized in the subsequent analysis. Here, $\mathbf{M}_{n}(f)$ denotes the set of spatial modes belonging to the $n^{\Th}$ mode-group at frequency $f$, encompassing both orthogonal polarizations, and ${D_{n}(f)=|\mathbf{M}_{n}(f)|}$ is the corresponding number of orthogonal spatial paths.

\section{Modeling of SRS in SDM fibers}
\label{sec:iii}

In this section, we provide analytical models to evaluate the combined effects of \gls*{srs} and crosstalk.

\subsection{Raman amplification}

Consider the interaction between a signal allocated at frequency $f_{s}$ and a significantly stronger pump at $f_{p}$. The total signal power in the $n^{\Th}$ mode-group evolves as (further detailed in \appref{app:a})
\begin{equation}
  \begin{split}
    \frac{\der P_{s,n}(z)}{\der z}\hspace{-2pt}=\hspace{-2pt}&-\hspace{-2pt}\alpha_{n}(f_{s})P_{s,n}(z)\hspace{-2pt}\\
    &+\sum_{m\neq n}\hspace{-4pt}D_{n}\kappa_{n,m}(f_{s})\hspace{-2pt}\left[P_{s,m}(z)\hspace{-2pt}-\hspace{-2pt}\frac{D_{m}}{D_{n}}P_{s,n}(z)\right]\\
    &+\Gr(\Delta f)\sum_{m}\Aeff{m,n}^{-1}P_{p,m}(z)P_{s,n}(z),
  \end{split}
  \label{eq:peqs}
\end{equation}
where $n,m\in(1,N)$ are the mode-group indices, $\alpha_{n}(f)$ is the mode-group-averaged attenuation, $\kappa_{n,m}(f)$ is the linearly averaged crosstalk coefficient between mode-groups $n$ and $m$ (see \appref{app:a1}), and $P_{p,m}(z)$ is the total pump power in mode-group $m$ at distance $z$. For simplicity, we neglect the frequency dependence of the effective area and mode count, as these values vary weakly with frequency. The parameter $\Gr(\Delta f)$, with ${\Delta f=|f_{p}-f_{s}|}$, is the Raman gain efficiency of the irradiated light, which depends on the fiber material properties (see~\cite{hellwarth1975origin} for an in-depth discussion).

Equation~\eqref{eq:peqs} can be written in vector form as
\begin{equation}
  \frac{\der \vec{P}_{s}(z)}{\der z}=\left\{-\boldsymbol{\alpha}_{s}+\mathbf{K}_{s}+\diag\left[\Gr\AeffMat\vec{P}_{p}(z)\right]\right\}\vec{P}_{s}(z),
  \label{eq:peqsmat}
\end{equation}
omitting the $\Delta f$ dependence of $\Gr$ for brevity, and where $\diag(\cdot)$ is the diagonal operator. The individual terms of the matrix $\AeffMat$ are the inverses of the mode-group-averaged self‐ and cross‐effective areas, and $\vec{P}_{p}(z)$ contains the pump mode‐group powers at distance $z$. Finally, the matrices $\boldsymbol{\alpha}_{s}$ and $\mathbf{K}_{s}$ contain the attenuation and crosstalk coefficients
\begin{equation}
  \begin{split}
    \boldsymbol{\alpha}&=\diag\left(
    \begin{bmatrix}
      \alpha_{1} & \alpha_{2} & \cdots & \alpha_{N}
    \end{bmatrix}
    \right),\\
    \mathbf{K}&=\setlength\arraycolsep{1pt}
    \begin{bmatrix}
      -\hspace{-6pt}\sum\limits_{m\neq 1}\hspace{-5pt}D_{m}\kappa_{1,m} & D_{1}\kappa_{1,2} & \cdots & D_{1}\kappa_{1,N}\\
      D_{2}\kappa_{2,1} & -\hspace{-6pt}\sum\limits_{m\neq 2}\hspace{-5pt}D_{m}\kappa_{2,m} & \cdots & D_{2}\kappa_{2,N}\\
      \vdots & \vdots & \ddots & \vdots \\
      D_{N}\kappa_{N,1} & D_{N}\kappa_{N,2} & \cdots & -\hspace{-6pt}\sum\limits_{m\neq N}\hspace{-7pt}D_{m}\kappa_{N,m}
    \end{bmatrix}.
  \end{split}
\end{equation}

Neglecting pump depletion, the pump power profiles can be solved analytically. The solution is distinct for co- and counter-propagation scenarios, respectively labeled ``\textit{Fwd}'' and ``\textit{Bwd}'' throughout this work, where for each we have
\begin{equation}
  \vec{P}_{p}(z)=\begin{cases}
    \exp\left[\left(-\boldsymbol{\alpha}_{p}+\mathbf{K}_{p}\right)z\right]\vec{P}_{p,\Tx},&\text{Fwd},\\
    \exp\left[\left(-\boldsymbol{\alpha}_{p}+\mathbf{K}_{p}\right)(L-z)\right]\vec{P}_{p,\Tx},&\text{Bwd},
  \end{cases}
\end{equation}
where $\boldsymbol{\alpha}_{p}$ and $\mathbf{K}_{p}$ are, respectively, the attenuation and crosstalk matrices at the pump frequency, and the subscript $(\cdot)_{p,\Tx}$ denotes the launched pump powers, at the fiber-end where the pump is located.

If the \gls*{sdm} link exhibits either sufficiently strong coupling, yielding a single degenerate mode-group, or negligible inter-mode-group interactions, scalar equations may be used, yielding expressions similar to those in \glspl*{smf}~\cite[Eq.~(1)]{sciullo2025characterization}. Nevertheless, the vector formulation presented here is general and applies regardless of coupling regime.

A exact closed-form solution for Eq.~\eqref{eq:peqs} is unknown. However, as shown in \appref{app:b}, a valid approximation is given by the first term of the Magnus expansion
\begin{equation}
  \begin{split}
    \vec{P}_{s}(z)\approx&\exp\left[\boldsymbol{\Omega}_{1}(z,0)\right]\vec{P}_{s}(0),\\
    \boldsymbol{\Omega}_{1}(z,0)=&\underbrace{\left(-\boldsymbol{\alpha}_{s}\hspace{-2pt}+\hspace{-2pt}\mathbf{K}_{s}\right)z}_{\text{Loss + crosstalk}}+\underbrace{\diag\left[\Gr\AeffMat\LeffMat(z)\vec{P}_{p,\Tx}\right]}_{\text{Raman amp. gain}},
  \end{split}
  \label{eq:pcf1}
\end{equation}
where $\LeffMat(z)$ is the matrix whose $(n,m)$ entry corresponds to the effective length of pump power launched in mode-group $m$ and measured in mode-group $n$, given by
\begin{equation}
  \LeffMat(z)\hspace{-3pt}=\hspace{-3pt}\begin{cases}
    \hspace{-2pt}\left[\mathbf{I}-e^{\left(-\boldsymbol{\alpha}_{p}+\mathbf{K}_{p}\right)z}\right]\left(\boldsymbol{\alpha}_{p}-\mathbf{K}_{p}\right)^{-1},&\text{Fwd},\\
    \hspace{-2pt}\left[e^{\left(-\boldsymbol{\alpha}_{p}+\mathbf{K}_{p}\right)(L-z)}\hspace{-2pt}-\hspace{-2pt}e^{\left(-\boldsymbol{\alpha}_{p}+\mathbf{K}_{p}\right)L}\right]\hspace{-3pt}\left(\boldsymbol{\alpha}_{p}\hspace{-2pt}-\hspace{-2pt}\mathbf{K}_{p}\right)^{\hspace{-2pt}-\hspace{-1pt}1}\hspace{-6pt},\hspace{-5pt}&\text{Bwd}.
  \end{cases}
\end{equation}

In some scenarios the first-order approximation is insufficient. It assumes that linear and non-linear effects commute, but high Raman gain may reshape the signal powers sufficiently to affect crosstalk evolution, and strong crosstalk may influence the experienced Raman gain. Therefore, \gls*{sdm} setups with intermediate\footnote{If the signals experience strong coupling, they may be treated as a single mode-group with equal loss, coupling, and gain. Otherwise, the first‐order solution can deviate noticeably, as shown in \appref{app:b}.} coupling and significant Raman gain may result in deviations from the first-order solution.

A more accurate closed‐form approximation is obtained by sectioning the fiber span in $V$ parts (see \appref{app:b1})
\begin{equation}
  \begin{split}
    \vec{P}_{s}(z)&\approx\left\{\prod_{\nu=V}^{1}\exp\left[\boldsymbol{\Omega}_{1}(z_{\nu},z_{\nu-1})\right]\right\}\vec{P}_{s}(0),\\
    \boldsymbol{\Omega}_{1}(z_{\nu},z_{\nu-1})&=\boldsymbol{\Omega}_{1}(z_{\nu},0)-\boldsymbol{\Omega}_{1}(z_{\nu-1},0),
  \end{split}
  \label{eq:cfsplit}
\end{equation}
where $z_{0}=0$ and $z_{0}<z_{1}<\cdots<z_{V}=z$. Increasing the section count $V$ reduces the approximation error at the cost of complexity. The required number of sections depends on the commutation error between the crosstalk and Raman‐gain matrices (see \appref{app:b}).

For bidirectional pumping with negligible pump-to-pump interactions, expanding~\eqref{eq:pcf1} yields
\begin{equation}
  \begin{split}
    \boldsymbol{\Omega}^{\text{Bi}}_{1}(z,0)=&\left(-\boldsymbol{\alpha}_{s}\hspace{-2pt}+\hspace{-2pt}\mathbf{K}_{s}\right)z\\
    &+\diag\{\Gr\AeffMat[\underbrace{\LeffMat^{F}(z)\vec{P}^{F}_{p,\Tx}}_{\text{Fwd}}+\underbrace{\LeffMat^{B}(z)\vec{P}^{B}_{p,\Tx}}_{\text{Bwd}}]\},
  \end{split}
  \label{eq:pcfbi}
\end{equation}
and the same reasoning applies for any number of pumps, as long as pump-to-pump interactions remain weak:
\begin{equation}
  \begin{split}
    \boldsymbol{\Omega}^{\Sigma}_{1}(z,0)=&\left(-\boldsymbol{\alpha}_{s}\hspace{-2pt}+\hspace{-2pt}\mathbf{K}_{s}\right)z\\
    &+\hspace{-2pt}\diag\hspace{-3pt}\left[\hspace{-1pt}\AeffMat\sum_{i}\Gr(f_{p,i}\hspace{-2pt}-\hspace{-2pt}f_{s})\LeffMat^{(i)}(z)\vec{P}^{(i)}_{p,\Tx}\right].
  \end{split}
\end{equation}

\subsection{On-off gain matrix}

It is useful to define the on-off gain of a given amplification setup. As significant levels of inter-mode-group Raman gain and crosstalk may occur, we define the on-off gain as a matrix $\mathbf{G}(z)$. Each entry $G_{n,m}(z)$ corresponds to the on-off gain experienced by a signal launched in mode-group $m$ and received in mode-group $n$ at distance $z$. Due to crosstalk, a signal spreads across multiple modes at the receiver, and each portion can experience distinct levels of Raman gain. As a result, a matrix-based formulation is unavoidable.

From the previous equations, the on-off gain matrix can be estimated as
\begin{equation}
  \mathbf{G}(z)=\left[\prod_{k=V}^{1} e^{\boldsymbol{\Omega}_{1}(z_{k},z_{k-1})}\right] \oslash e^{\left(-\boldsymbol{\alpha}_{s}+\mathbf{K}_{s}\right)z},
\end{equation}
where $\oslash$ is the elementwise division operator.

To compare different Raman amplification schemes, we define two figures of merit: the average on-off Raman gain $\mu_{\mathbf{G}}$ and the \gls*{mdg} deviation $\sigma_{\mathbf{G}}$. The average on-off Raman gain represents the expected gain of a signal launched in a randomly chosen supported mode among all mode-groups and is quantified as
\begin{equation}
  \mu_{\mathbf{G}}=\frac{\sum_{n,m\in(1,N)}D_{n}G_{n,m}(z)}{N\sum_{n\in(1,N)}D_{n}}.
  \label{eq:mug}
\end{equation}

The \gls*{mdg} is an important performance metric in \gls*{sdm} transmission, as deviations in modal gains are known to degrade achievable information rates~\cite{ho2011mode,mello2020impact} and reduce the effectiveness of equalization techniques~\cite{zischler2025sdm}. The \gls*{mdg} is evaluated in decibels, as defined in previous literature~\cite{ho2011mode}, and can be calculated by
\begin{equation}
  \sigma^{\dB}_{\mathbf{G}}=\sqrt{\frac{\left[\sum_{n,m\in(1,N)}D_{n}G^{\dB}_{n,m}(z)-\mu^{\dB}_{\mathbf{G}}\right]^{2}}{\left(N\sum_{n\in(1,N)}D_{n}\right)-1}},
\end{equation}
where, $G^{\dB}_{n,m}(z)$ is the on-off Raman gain in decibels, and $\mu^{\dB}_{\mathbf{G}}$ is the weighted average gain obtained by applying~\eqref{eq:mug} using the logarithmic-domain gains $G^{\dB}_{n,m}$, rather than its linear-scale equivalent\footnote{Mode-dependent loss and gain accumulates multiplicatively. Therefore, \gls*{mdg}-related penalties are quantified with the log-domain metric~\cite[Section~4]{ho2011mode}. The correct computation of the standard deviation requires averaging in the same domain, resulting in distinct values for $\mu_{\mathbf{G}}$ and $\mu^{\dB}_{\mathbf{G}}$. Their analytical relationship is provided in better detail in~\cite[Appendix~B]{zischler2025analytic}.}.

\subsection{Raman ASE noise}

In the absence of incident signal photons, phonons in the excited state decay spontaneously into random states, resulting in \gls*{sprs}. The \gls*{psd} of \gls*{sprs} photons in a given propagation direction consists of \gls*{srs}-amplified vacuum fluctuations and thermally excited phonons, which follow Bose-Einstein statistics. Similar to conventional optical amplification, this process generates \gls*{ase} noise, when scattered photons are amplified by \gls*{srs}. A study of \gls*{sprs} in \gls*{sdm} fibers is provided in~\cite{zischler2025spontaneous}.

The accumulated \gls*{sprs}-induced \gls*{ase} noise \gls*{psd} $\vec{S}^{\eta}_{s}(z)$ within mode-groups at frequency $f_{s}$ evolves in vector form as~\cite[Eq.~(6)]{bromage2004raman},~\cite[Eq.~(7)]{zischler2025spontaneous}
\begin{equation}
  \pm\frac{\der \vec{S}^{\eta}_{s}(z)}{\der z}\hspace{-3pt}=\hspace{-3pt}\Big\{\hspace{-3pt}-\hspace{-2pt}\boldsymbol{\alpha}_{s}\hspace{-3pt}+\hspace{-3pt}\mathbf{K}_{s}\hspace{-3pt}+\hspace{-2pt}\underbrace{\diag\hspace{-2pt}\left[\Gr\AeffMat\vec{P}_{p}(z)\right]}_{\text{Raman amp. gain}}\hspace{-3pt}\Big\}\hspace{-1pt}\vec{S}^{\eta}_{s}(z)\hspace{-2pt}+\hspace{-2pt}\underbrace{\boldsymbol{\eta}\vec{P}_{p}(z)}_{\text{SpRS}},
  \label{eq:sprs}
\end{equation}
where the elements of the matrix $\boldsymbol{\eta}$ are the mode-group-averaged \gls*{sprs} spectral efficiency values $\eta_{n,m}(\Delta f)$, as described in~\cite{zischler2025spontaneous}. Analogous to the Raman gain, the \gls*{sprs} efficiency can be separated into a material-dependent term $\eta_{\mathrm{R}}$ and a spatial term, such that ${\boldsymbol{\eta}(\Delta f)=\eta_{\R}(\Delta f)\AeffMat}$.

The \gls*{osnr} at the $n^{\Th}$ mode-group in the presence of \gls*{dra} at any propagation distance $z$ is then given by
\begin{equation}
  \OSNR_{n}(z)=\frac{P_{s,n}(z)}{S^{\eta}_{s,n}(z)+G_{n}(z)\sigma^{2}_{\text{Noise},n}},
\end{equation}
where, $B_{\rf}$ is the signal bandwidth and $\sigma^{2}_{\text{Noise},n}$ denotes the noise variance at the input of the fiber span, allowing a straightforward iterative evaluation in multi-span scenarios.

\subsection{\Glsxtrlong*{isrs}}

\Glsxtrlong*{isrs} is particularly significant in wideband transmission, where interaction between many sufficiently spaced channels results in a spectral tilt. In contrast to \gls*{dra}, depletion must be taken into account, and the Raman efficiency can be approximated by a linear slope $(\Gr(\Delta f)\approx \Cr\,\Delta f)$ for transmission windows below 15~THz~\cite{chraplyvy1984optical}.

Considering a continuous power spectra between $\fmin$ and $\fmax$, and extending \eqref{eq:peqs} following \cite[Eq.~(2)]{zirngibl1998analytical}, the \gls*{psd} of the $n^{\Th}$ mode-group $S_{n}(f,z)$ evolves according to
\begin{equation}
  \begin{split}
    \frac{\partial S_{n}(f,z)}{\partial z}\hspace{-2pt}=\hspace{-2pt}&-\hspace{-2pt}\alpha_{n} S_{n}(f,z)\hspace{-2pt}\\
    &+\hspace{-2pt}\sum_{m\neq n}\hspace{-4pt}D_{n}\kappa_{n,m}\hspace{-2pt}\left[S_{m}(f,z)\hspace{-2pt}-\hspace{-2pt}\frac{D_{m}}{D_{n}}S_{n}(f,z)\right]\\
    &+\hspace{-2pt}\Cr\hspace{-4pt}\int_{\fmin}^{\fmax}\hspace{-20pt}(f'-f)\hspace{-2pt}\sum_{m}\hspace{-2pt}\Aeff{m,n}^{-1}S_{m}(f',z)S_{n}(f,z)\der f'\hspace{-2pt},
  \end{split}
  \label{eq:pisrs}
\end{equation}
where we assume flat attenuation and a flat coupling profile within the transmission window of each mode-group, and represented in the following vector representation
\begin{equation}
  \frac{\partial \vec{S}(f,z)}{\partial z}\hspace{-2pt}=\hspace{-2pt}\left\{\hspace{-2pt}-\boldsymbol{\alpha}\hspace{-2pt}+\hspace{-2pt}\mathbf{K}\hspace{-2pt}+\hspace{-2pt}\Cr\hspace{-4pt}\int_{\fmin}^{\fmax}\hspace{-20pt}(f'\hspace{-4pt}-\hspace{-3pt}f)\diag\hspace{-2pt}\left[\AeffMat\vec{S}(f',z)\right]\hspace{-2pt}\right\}\hspace{-2pt}\vec{S}(f,z),
\end{equation}

Because \gls*{isrs} within conventional transmission bands does not introduce significant \gls*{srs} gain or loss, we may assume crosstalk and Raman effects as independent. This assumption also implies that \gls*{isrs} dynamics have only a minor influence on the total mode-group power, defined as
\begin{equation}
 \vec{P}_{\Total}(z)=\int_{\fmin}^{\fmax}\vec{S}(f',z)\der f'=\exp\left[\left(-\boldsymbol{\alpha}+\mathbf{K}\right)z\right]\vec{P}_{\Total}(0).
 \label{eq:ptotal}
\end{equation}

The previous considerations, allow us to adopt the simplified solution obtained from the first term of the Magnus expansion, resulting in
\begin{equation}
  \begin{split}
    \vec{S}(f,z)\hspace{-2pt}\approx&\exp\left[\boldsymbol{\tilde{\Omega}}_{1}(f,z)\right]\vec{S}(f,0),\\
    \boldsymbol{\tilde{\Omega}}_{1}(f,z)\hspace{-2pt}=&\hspace{-2pt}\underbrace{\left(-\boldsymbol{\alpha}\hspace{-2pt}+\hspace{-2pt}\mathbf{K}\hspace{-1pt}\right)\hspace{-2pt}z}_{\text{Loss + crosstalk}}\hspace{-2pt}+\hspace{-1pt}\underbrace{\Cr\diag(\Fr\hspace{-2pt}-\hspace{-2pt}f)\diag\hspace{-2pt}\left[\AeffMat\LeffMat(z)\vec{P}_{\Total}(0)\right]}_{\text{ISRS relative gain/loss}},
  \end{split}
  \label{eq:cfisrs}
  \raisetag{24pt}
\end{equation}
where the elements of ${\Fr=[\Frm{1},\Frm{2},\ldots,\Frm{D}]}$ are the zero-\gls*{isrs}-induced-tilt frequencies of the mode-groups. These frequencies values bear similarity to \cite[Eq.~(10)]{zischler2025closed}, satisfying \gls*{srs} energy conservation, and are given by (derived in \appref{app:c})
\begin{equation}
  \Frm{n}=\frac{-1}{\Cr P_{\eff,n}(z)}\ln\left[\int_{\fmin}^{\fmax}\frac{S_{n}(f,0)}{P_{\Total,n}(0)}e^{-\Cr f P_{\eff,n}(z)}\right],
  \label{eq:fr}
\end{equation}
where $P_{\eff,n}(z)$ is utilized for simplicity as the elements of
\begin{equation}
  \vec{P}_{\eff}(z)=\AeffMat\LeffMat(z)\vec{P}_{\Total}(0).
  \label{eq:peff}
\end{equation}

Equation~\eqref{eq:fr} admits a closed-form expression when discrete channels are considered, in which case the frequency integral is replaced by a summation over the channels.

\section{Analytical Evaluation}
\label{sec:iv}

\begin{table}[!t]
    \centering
    \begin{tabular}{|c|c|} 
         \hline
         \textbf{\makebox[2cm][c]{Parameter}} & \textbf{\makebox[2cm][c]{Value}} \\\hline
         Effective SRS peak $\GR$ (@ 15~THz)\hspace{-2pt} & \SI{50}{\micro\metre\squared\per\watt\per\kilo\metre} \\
         Effective SpRS eff. peak $\eta_{\R}$ & \SI{7.5e-9}{\micro\metre\squared\per\kilo\metre\per\giga\hertz}\hspace{-2pt} \\
         Signal freq. attenuation coef. $\alpha_{s}$ & 0.2 dB/km \\
         Pump freq. attenuation coef. $\alpha_{p}$ & 0.25 dB/km \\
         Span length $L_{s}$ & 50 km \\
         Simulation steps & $10^{4}$ \\
         Section count $V$ & 20 \\
         \hline
    \end{tabular}
    \caption{Parameters for analytical and simulation results.}
    \label{tab:parameters}
\end{table}

\begin{figure}[!ht]
    \centering
    \includegraphics{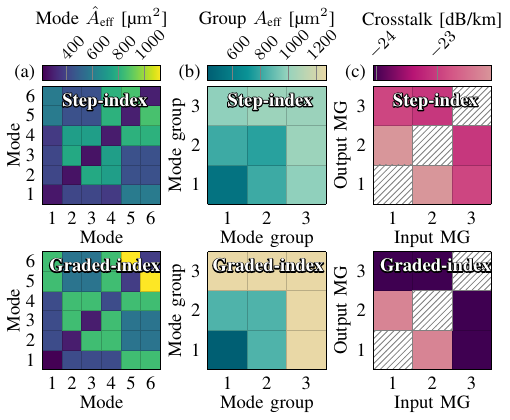}
    \caption{Mode- and mode-group-dependent parameters of the considered \gls*{si-mmf} and \gls*{gi-mmf}. (a)~Self- and cross-effective mode areas, (b)~mode-group-averaged values, and (c)~coupling coefficients between distinct mode-groups.}
    \label{fig:MmfParam}
\end{figure}

\begin{figure*}[!t]
    \centering
    \includegraphics{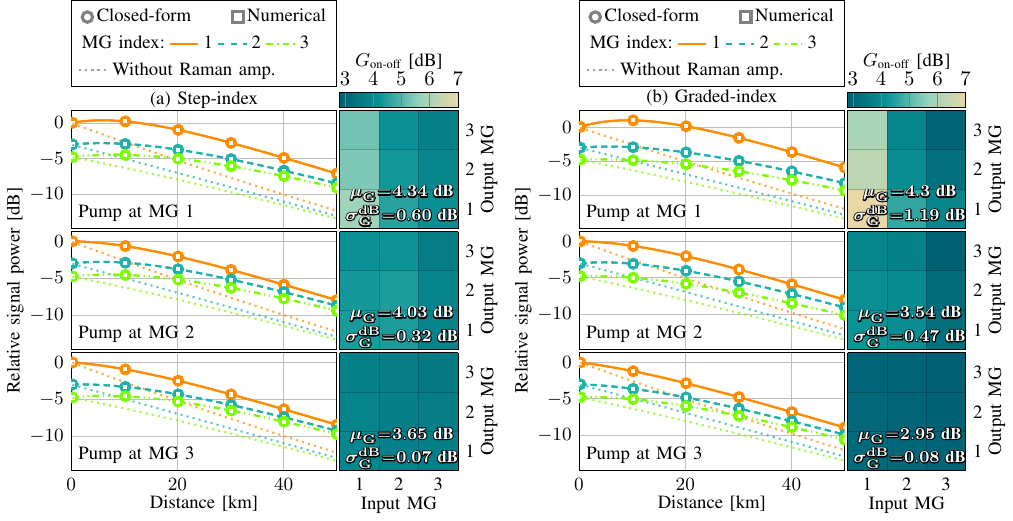}
    \caption{Longitudinal profiles of the mode-group-averaged signal powers, normalized to the fundamental mode, with a Raman pump applied to the specified mode-group for (a) a \gls*{si-mmf} and (b) a \gls*{gi-mmf}. The circle and square markers correspond to the closed-form and numerical solutions, respectively. Distinct line styles denote the mode-group index. The dotted lines show the power profiles in the absence of the Raman pump. To the right of the power profiles, the on-off gain matrix for each scenario is shown, with the average gain ($\mu_{\mathbf{G}}$) and \gls*{mdg} ($\sigma^{\dB}_{\mathbf{G}}$) displayed alongside the plots.}
    \label{fig:MmfDra}
\end{figure*}

Utilizing the analytical framework developed in the previous section, we evaluate the Raman effect under representative scenarios. Unless otherwise specified, all analytical and simulation results utilize the parameters listed in Tab.~\ref{tab:parameters}. Mode-group-dependent parameters are shown in Fig.~\ref{fig:MmfParam} for the respective \gls*{mmf} types.

The chosen effective \gls*{srs} and \gls*{sprs} efficiencies, listed in Tab.~\ref{tab:parameters}, correspond to \SI{0.3125}{\per\watt\per\kilo\metre} and \SI{4.7e-11}{\per\kilo\metre\per\giga\hertz}, respectively, for a typical \SI{80}{\micro\metre\squared} effective-area \gls*{smf} (\SI{160}{\micro\metre\squared} polarization-averaged $A_{\eff}$), reflecting practical experimental parameters~\cite{rottwitt2003scaling, eraerds2010quantum, sciullo2025characterization, zischler2025spontaneous}. Effective area values, shown in Fig.~\ref{fig:MmfParam}(a,b), are calculated from the respective fundamental lateral field profile functions~\cite[Chapter 14]{agrawal2007nonlinear}. The crosstalk values, shown in Fig.~\ref{fig:MmfParam}(c), are chosen reflecting experimental measurements~\cite{rodigheri2025experimental, rademacher2021peta}, scaled by the effective area.

\subsection{\Glsxtrlong*{dra}}

Figure~\ref{fig:MmfDra} shows the longitudinal mode-group-averaged signal power profiles under a co-propagating 30~dBm pump for \gls*{dra}. The signal launch powers are arbitrarily set as constant per mode-group, resulting in crosstalk from the lowest-order mode-group to the highest. Closed-form curves are calculated from Eq.~\eqref{eq:cfsplit} and compared against numerical integrations of Eq.~\eqref{eq:peqs}, solved with a Runge-Kutta method. As shown, the closed-form and numerical solutions overlap closely, exhibiting negligible deviation for the considered parameters.

In the \gls*{si-mmf}, the higher modal overlap, shown in Fig.~\ref{fig:MmfParam}(a,b), results in \gls*{srs} gain being more evenly spread among the mode-groups compared with the \gls*{gi-mmf}, at the cost of increased modal dispersion~\cite{carpenter2012degenerate}. This behavior is quantitatively reflected in the respective \gls*{mdg} values, which are consistently larger for the \gls*{gi-mmf} under identical configurations.

Placing the pump in the lowest-order mode-group yields higher \gls*{srs} gain, most notably in the fundamental mode-group of the \gls*{gi-mmf}. However, this configuration also results in increased \gls*{mdg}. In the \gls*{si-mmf}, increasing the pump mode-group order produces only a minor reduction in average gain, in contrast to the more significant decrease observed in the \glspl*{gi-mmf}.

In scenarios with negligible crosstalk, where mode-groups can be independently decoded at the fiber output and inter-mode-group \gls*{mdg} does not significantly degrade transmission rates, placing the Raman pump in the fundamental mode may be preferable to maximize gain. However, excluding mode-groups during filtering can lower receiver complexity at the expense of capacity penalties~\cite{ospina2023experimental}. When mode-group \gls*{mimo} processing is employed, a more balanced equalization profile may be desirable, in which case placing the pump in the highest-order mode-group may reduce \gls*{snr} oscillations.

\begin{figure}[!t]
    \centering
    \includegraphics{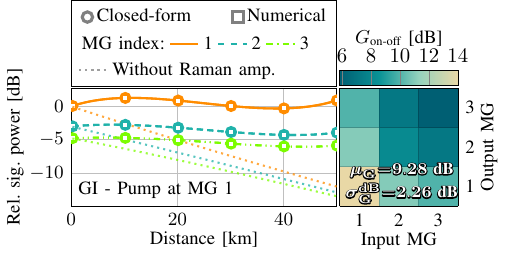}
    \caption{Longitudinal profiles of mode-group-averaged powers, normalized to the fundamental mode, under bidirectional \gls*{dra} in the \gls*{gi-mmf}. The corresponding on-off gain matrix is shown on the right.}
    \label{fig:Bidir}
\end{figure}

Figure~\ref{fig:Bidir} shows the longitudinal mode-group-averaged signal powers in a bidirectional amplification scenario, where both pumps are launched with 30~dBm at the fundamental mode-group. The closed-form curves are expanded to account for the two pump directions, as given by Eq.~\eqref{eq:pcfbi}, and again show excellent agreement with the numerical solutions. Compared with the co-propagating configurations in Fig.~\ref{fig:MmfDra}, where the highest gains occur for signals launched in the first mode-group, the bidirectional case in Fig.~\ref{fig:Bidir} yields a symmetric on-off gain matrix due to the counter-propagating pump amplifying primarily the fundamental mode-group near the fiber end.

\begin{figure}[!t]
    \centering
    \includegraphics{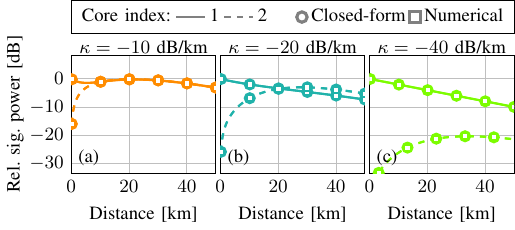}
    \caption{Longitudinal power profile of a signal launched in core~1 of a 2-core \gls*{mcf}, with a 30~dBm co-propagating Raman pump placed in the opposite core. Results are shown for representative (a) strong, (b) intermediate, and (c) weak coupling regimes. Line styles indicate the respective core index.}
    \label{fig:McfDra}
\end{figure}

Figure~\ref{fig:McfDra} shows a \gls*{dra} scenario in which the signal is launched in a dedicated core of a 2-core \gls*{mcf}, while a 30~dBm co-propagating pump is injected into the adjacent core. The fundamental mode effective areas of both cores are  defined as \SI{80}{\micro\metre\squared}.

The coupling intensity in \glspl*{mcf} can be coarsely tuned in the fiber design by adjusting the core separation and by incorporating low refractive index trenches~\cite{hayashi2011design,saitoh2016multicore}. The evolution of \gls*{srs} can differ for distinct coupling regimes, as seen in~\cite{sciullo2025characterization}. In Fig.~\ref{fig:McfDra}(a), under strong coupling, the signal and pump powers equalize after a few kilometers, where the two cores effectively behave as a single mode-group and experience identical gain. In the intermediate coupling regime, shown in Fig.~\ref{fig:McfDra}(b), the signal portion leaking into the adjacent core becomes stronger than the portion of the signal left in the launch core. This can occur because the \gls*{srs} gain is concentrated near the fiber input and rapidly saturates, as implied by the effective-length term of~\eqref{eq:pcf1}, whereas crosstalk grows approximately linearly with distance. Consequently, the leaked signals may be significantly amplified before crosstalk equalizes the core powers. In the weak coupling regime, shown in Fig.~\ref{fig:McfDra}(c), only a negligible fraction of the signal transfers to the adjacent core. Since the pump remains largely confined to that adjacent core, the original signal experiences almost no gain, where only the small leaked portion benefits from amplification.

\subsection{Raman ASE noise}

\begin{figure}[!t]
    \centering
    \includegraphics{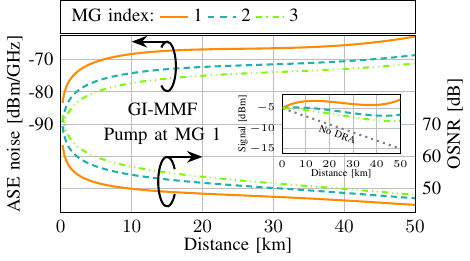}
    \caption{Longitudinal profiles of mode-group-averaged \gls*{ase} and \gls*{osnr}. Inset shows the corresponding longitudinal mode-group-averaged signal power profiles.}
    \label{fig:Sprs}
\end{figure}

In Fig.~\ref{fig:Sprs}, we plot the evolution of the accumulated mode-group-averaged \gls*{ase} in a \gls*{gi-mmf} with co- and counter-propagating 30~dBm pumps launched in the fundamental mode-group. Alongside the \gls*{ase} curves, we plot the resulting \gls*{osnr} for 32~GHz signals launched at -5~dBm per mode. The noise curves are obtained numerically from Eq.~\eqref{eq:sprs}.

As observed in Fig.~\ref{fig:Sprs}, noise accumulates at different rates across the mode-groups, since both \gls*{sprs} and \gls*{srs} scale with the effective area. This mode-dependent noise, analogous to \gls*{mdg}, degrades achievable capacity in \gls*{sdm} links, resulting in mode-dependent \gls*{snr}, as discussed in the literature~\cite{ho2011mode, antonelli2015modeling, mello2020impact}.

\begin{figure}[!t]
    \centering
    \includegraphics{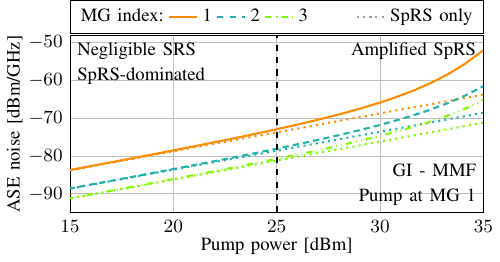}
    \caption{Mode-group-averaged amplification noise versus counter-propagating pump power allocated to the fundamental mode-group for the \gls*{gi-mmf}.}
    \label{fig:SprsPower}
\end{figure}

Amplification noise is proportional to the \gls*{srs} gain, which increases with pump power. The \gls*{psd} of \gls*{sprs}-generated photons grows linearly with pump power, while the \gls*{srs} gain grows exponentially, as shown in Eq.~\eqref{eq:pcf1}. Consequently, different noise effects dominate in different pump power regions, where the crossover point between these regimes depends on the fiber design.

In Fig.~\ref{fig:SprsPower}, we plot the \gls*{ase} noise \gls*{psd} at 50~km for a counter-propagating pump launched in the fundamental mode-group. For low pump powers, unamplified \gls*{sprs} dominates. As the pump power increases, \gls*{srs} becomes dominant, and \gls*{sprs} is further amplified as \gls*{ase}.

\subsection{\Glsxtrlong*{isrs}}

\begin{figure}[!t]
    \centering
    \includegraphics{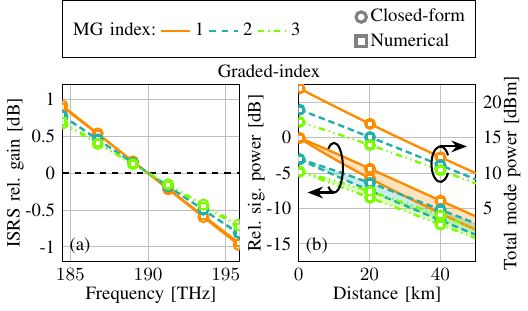}
    \caption{(a) Mode-group-averaged \gls*{isrs} relative gain versus frequency at 50~km. (b) Longitudinal signal power profiles, normalized to the fundamental mode, alongside the total mode-group powers. Signal power curves are shown for the edge frequencies, with the filled regions indicating the intermediate channels.}
    \label{fig:Isrs}
\end{figure}

Figure~\ref{fig:Isrs} shows the \gls*{isrs} evolution in a \gls*{gi-mmf} where C+L band signals are launched with a constant mode-group power of 22~dBm. The closed-form curves consider 100~GHz-spaced discrete channels, computed using Eq.~\eqref{eq:cfisrs}, and are plotted against numerical solutions of Eq.~\eqref{eq:pisrs}, showing great agreement.

Power transfer from L- to C-band channels results in a distorted received spectral power profile. As observed in Fig.~\ref{fig:Isrs}(a), equivalent to \gls*{dra} gain values, the lowest mode-group experiences the highest \gls*{isrs} tilt, in agreement with experimental observations~\cite{schneck2025experimental} and previous analytical work~\cite{zischler2025evaluation}. Figure~\ref{fig:Isrs}(b) shows mode powers converging due to crosstalk, therefore, reducing \gls*{isrs} deviations between mode-groups.

\section{Fiber characterization}
\label{sec:v}

As the Raman response depends on the fiber doping and mode fields, each fiber can exhibit distinct \gls*{srs} profiles. Accurate characterization of the fiber's Raman profile is therefore necessary for proper modeling of the link. Nevertheless, the combined interactions of coupling and inter-modal \gls*{srs} makes it difficult to perform straightforward measurements.

In this section, we assume the fiber to be a homogeneous medium along its longitudinal axis. This excludes lumped losses and coupling introduced by connectors, multiplexers, and other passive optical devices. These devices typically have a linear response that can often be characterized independently from the fiber link.

Given a set of $N$ transmitted power vectors $\vec{P}^{M}_{i}(0)$ measured or estimated at the fiber input, along with the corresponding received vectors at an accessible distance $L$ ($\vec{P}^{M}_{i}(L)$) a transmitted matrix $\mathbf{\bar{P}}(0)$ can be formed by concatenating the input vectors as ${\mathbf{\bar{P}}(0) = [\vec{P}^{M}_{i_{1}}(0)\,|\,\vec{P}^{M}_{i_{2}}(0)\,|\,\cdots\,|\,\vec{P}^{M}_{i_{N}}(0)]}$. A received matrix $\mathbf{\bar{P}}(L)$ can be formed analogously.

\subsection{Crosstalk measurements}

We briefly review the procedure for measuring the crosstalk matrix, which is required to extract the Raman response from measured signals in fibers with non-negligible mode coupling. In the absence of Raman effects, the attenuation and coupling matrices can be estimated from
\begin{equation}
  \boldsymbol{\alpha}-\mathbf{K}=\frac{-1}{L}\ln\left[\mathbf{\bar{P}}(L)\mathbf{\bar{P}}^{-1}(0)\right]
\end{equation}
where, when applied to matrices, $\ln(\cdot)$ defines the matrix logarithm operator.

\subsection{Pump gain measurements}

With pump gain measurements, the \gls*{srs} profile of a given fiber can be derived. For such measurements, a probe signal at a frequency lower than the pump ($f_{s} < f_{p}$) is launched and propagated along the fiber link. To obtain accurate estimates, certain conditions must be satisfied. The probe signal must be significantly weaker than the pump so that pump depletion can be neglected, and the \gls*{srs} gain or mode-group coupling values must remain within the regime where Eq.~\eqref{eq:pcf1} provides a valid approximation. Since \gls*{sprs} or \gls*{ase} can dominate over the probe power, either the probe must be made significantly stronger than the \gls*{ase}, or the \gls*{ase} must be subtracted from the measurements. The latter can be achieved by measuring the \gls*{ase} at the probe frequency with the probe signal turned off~\cite{sciullo2025characterization}.

By fixing the pump powers within the mode-groups and measuring the received probe power for given launch profiles $\mathbf{\bar{P}}_{s,\Tx}$, with the pump on $\mathbf{\bar{P}}^{\text{ON}}_{s,\Rx}$ and with the pump off $\mathbf{\bar{P}}^{\text{OFF}}_{s,\Rx}$, the \gls*{srs} efficiency can be estimated from
\begin{equation}
  \begin{split}
    \Gr=&\left[\ln\left(\mathbf{\bar{P}}^{\text{ON}}_{s,\Rx}\mathbf{\bar{P}}^{-1}_{s,\Tx}\right)-\ln\left(\mathbf{\bar{P}}^{\text{OFF}}_{s,\Rx}\mathbf{\bar{P}}^{-1}_{s,\Tx}\right)\right]\\
        &\times \diag\left[\AeffMat\LeffMat(L)\vec{P}_{p,\Tx}\right]_{n,n}^{-1}.
  \end{split}
\end{equation}
for any arbitrary mode-group index $n$. The resulting \gls*{srs} efficiencies should ideally be equal for all mode-group measurements. A significant deviation from this behavior indicates that the system configuration may be outside the regime where Eq.~\eqref{eq:pcf1} provides an accurate approximation.

\subsection{ISRS-induced tilt}

With \gls*{isrs} measurements, it is possible to estimate the linear slope of the Raman gain efficiency. In general, a closed-form solution cannot be obtained for arbitrarily loaded spectra, since the relative \gls*{isrs} gain is proportional to the total mode-group powers, which prevents the concatenation of different measurements. Nevertheless, if the modes are equally loaded so that crosstalk can be considered negligible\footnote{This assumption was previously adopted in~\cite{schneck2025experimental,zischler2025evaluation}, showing good agreement with experiments.}, the tilt (quantified as the difference between the spectral edges) in the $n^{\Th}$ mode-group is given by~\cite[Eq.~(4)]{zischler2025evaluation}
\begin{equation}
  \text{Tilt}^{\dB}_{n}=\frac{10}{\ln(10)}\Cr(\fmin-\fmax)\diag\left[\AeffMat\LeffMat(L)\vec{P}_{\Total}(0)\right]_{n,n},
\end{equation}
where, under this condition, the Raman gain slope can be estimated for any arbitrary mode-group $n$ as
\begin{equation}
  \Cr=\frac{\ln(10)\text{Tilt}^{\dB}_{n}}{10(\fmin-\fmax)}\diag\left[\AeffMat\LeffMat(L)\vec{P}_{\Total}(0)\right]^{-1}_{n,n}.
\end{equation}

\subsection{SpRS noise measurements}

Within the regime of negligible \gls*{srs} gain, Eq.~\eqref{eq:sprs} can be solved analytically as
\begin{equation}
  \vec{S}^{\eta}_{s}(z)=\begin{cases}
    \int_{0}^{z}e^{(-\boldsymbol{\alpha}_{s}+\mathbf{K}_{s})(z-z')}\boldsymbol{\eta}e^{(-\boldsymbol{\alpha}_{p}+\mathbf{K}_{p})z'}\vec{P}_{p,\Tx}\der z',&\text{Fwd},\\
    \int_{z}^{L}e^{(-\boldsymbol{\alpha}_{s}+\mathbf{K}_{s})(z'-z)}\boldsymbol{\eta}e^{(-\boldsymbol{\alpha}_{p}+\mathbf{K}_{p})z'}\vec{P}_{p,\Tx}\der z',&\text{Bwd},
  \end{cases}
\end{equation}
where the fiber \gls*{sprs} response can be estimated from any arbitrary mode-group as
\begin{equation}
  \eta_{\R}=\begin{cases}
    \frac{S^{\eta}_{s,n}(z)}{\left[\int_{0}^{z}e^{(-\boldsymbol{\alpha}_{s}+\mathbf{K}_{s})(z-z')}\AeffMat e^{(-\boldsymbol{\alpha}_{p}+\mathbf{K}_{p})z'}\vec{P}_{p,\Tx}\der z'\right]_{n}},&\text{Fwd},\\
    \frac{S^{\eta}_{s,n}(z)}{\left[\int_{z}^{L}e^{(-\boldsymbol{\alpha}_{s}+\mathbf{K}_{s})(z'-z)}\AeffMat e^{(-\boldsymbol{\alpha}_{p}+\mathbf{K}_{p})z'}\vec{P}_{p,\Tx}\der z'\right]_{n}},&\text{Bwd}.
  \end{cases}
\end{equation}

As seen in~\cite[Eq.~(8)]{zischler2025spontaneous}, the \gls*{sprs} efficiency can be inferred from the Raman gain efficiency. Likewise, the reverse is possible with
\begin{equation}
  \Gr(\Delta f)=\begin{cases}
    \eta_{\R}(\Delta f)\Big\{[1+\Phi(\Delta f)]hf_{s}\Big\}^{-1},& f_{s}<f_{p},\\
    \eta_{\R}(\Delta f)\Big[\Phi(\Delta f)hf_{s}\Big]^{-1},& f_{s}>f_{p},\\
  \end{cases}
\end{equation}
where $h$ is Planck's constant and $\Phi(\Delta f)$ is the occupancy factor of thermally excited phonons~\cite[Eq.~(25.35)]{ashcroft1976solid}.

\section{Conclusion}
\label{sec:vi}

The study of non-linear interactions in novel \gls*{sdm} fiber systems relies on the careful consideration of many parameters, such as the mode fields distributions and inter-modal crosstalk. In this work, a comprehensive assessment of the Raman effect in \gls*{sdm} transmission links is provided, complementing previous studies on Raman amplification and \gls*{isrs} with general closed-form expressions, showing excellent agreement to numerical simulations. From the analytical formulas, representative scenarios are evaluated, demonstrating how inter-mode-group \gls*{srs} and coupling regimes can affect \gls*{srs} gains and \gls*{ase} noise.

\appendices
\def\sectionautorefname{appendix}

\section{Mode-group-averaged coupled-power equations}
\label{app:a}

From coupled-power equations defined in the literature~\cite{gloge1972optical,marcuse1973coupled}, with the additional \gls*{srs} contribution~\cite[Eqs.~(19--20)]{antonelli2013raman}, the power $\hat{P}_{s,i}(z)$ at frequency $f_{s}$ in an individual mode $i$ evolves according to
\begin{equation}
  \begin{split}
    \frac{\der \hat{P}_{s,i}(z)}{\der z}\hspace{-2pt}=\hspace{-2pt}&-\hspace{-2pt}\hat{\alpha}_{i}(f_{s})\hat{P}_{s,i}(z)\hspace{-2pt}+\hspace{-4pt}\sum_{h\neq i}\hspace{-2pt}\hat{\kappa}_{i,h}(f_{s})\hspace{-2pt}\left[\hat{P}_{s,h}(z)\hspace{-2pt}-\hspace{-2pt}\hat{P}_{s,i}(z)\right]\\
    &+\Gr(\Delta f)\sum_{h}\Aeffm{i,h}^{-1}\hat{P}_{p,h}(z)\hat{P}_{s,i}(z).
  \end{split}
  \label{eq:a1}
\end{equation}
where the hat $\hat{\cdot}$ denotes per-mode quantities.

Within a given mode-group, the power is evenly distributed among modes due to strong crosstalk, so that ${\hat{P}_{s,i}(z)=P_{s,n}(z)/D_{n}(f_{s})}$, and the attenuation coefficients can be replaced by their mode-group average ${\hat{\alpha}_{i}(f) \rightarrow \alpha_{n}(f)}$.

\subsection{Coupling coefficient}
\label{app:a1}

If the $i^{\Th}$ and $h^{\Th}$ mode belong to the same $n^{\Th}$ mode-group, crosstalk can be neglected under the assumption of equal power ($\kappa_{i,h}(f)=0$, $i,h \in \mathbf{M}_{n}$). Considering only inter-group crosstalk, at an arbitrary frequency, the mode-group powers evolve according to
\begin{equation}
  \frac{\der P_{n}(z)}{\der z}=\sum_{m\neq n}D_{n}\kappa_{n,m}\left[P_{m}(z)-\frac{D_{m}}{D_{n}}P_{n}(z)\right],
  \label{eq:a2}
\end{equation}
where the mode-group-averaged crosstalk coefficient $\kappa_{n,m}$ is obtained from
\begin{equation}
  \kappa_{n,m}=\sum_{\substack{i\in\mathbf{M}_{n}\\h\in\mathbf{M}_{m}}}\frac{\hat{\kappa}_{i,h}}{D_{n}D_{m}}.
  \label{eq:a3}
\end{equation}

\subsection{SRS gain/loss}

Considering only \gls*{srs}, the mode-group power evolves according to~\cite{antonelli2013raman}
\begin{equation}
  \frac{\der P_{s,n}(z)}{\der z}=\Gr(\Delta f)\sum_{m}\hspace{-4pt}\sum_{\substack{i\in\mathbf{M}_{n}\\h\in\mathbf{M}_{m}}}\hspace{-4pt}\Aeffm{i,h}^{-1}\frac{P_{p,m}(z)}{D_{m}(f_{p})}\frac{P_{s,n}(z)}{D_{n}(f_{s})},
  \label{eq:a4}
\end{equation}
where, by using the mode-group-averaged cross-effective area defined in Eq.~\eqref{eq:aeffmg}, we obtain
\begin{equation}
  \frac{\der P_{s,n}(z)}{\der z}=\Gr(\Delta f)\sum_{m}\Aeff{n,m}^{-1}P_{p,m}(z)P_{n,p}(z).
  \label{eq:a5}
\end{equation}

By substituting Eqs.~\eqref{eq:a2} and~\eqref{eq:a5} into the coupled-power equations in~\eqref{eq:a1}, the system can be rewritten as~\eqref{eq:peqs}.

\section{Closed-form approximated solution}
\label{app:b}

As the system is described by a set of linear $z$-dependent ordinary differential equations, the mode-group powers at any arbitrary position $z_{\mu}$ can be expressed as a linear combination of the values at a second position $z_{\nu}$. Nevertheless, a general analytically exact solution for Eq.~\eqref{eq:peqs} can only be found in infinite series such as the Magnus expansion~\cite{magnus1954exponential},~\cite[Eqs.~(47--50)]{blanes2009magnus}
\begin{equation}
  \begin{split}
    \vec{P}_{s}(z)&=\exp\left[\sum_{k=1}^{\infty}\boldsymbol{\Omega}_{k}(z,0)\right]\vec{P}_{s}(0)\\
  \end{split}
\end{equation}
where the matrices $\boldsymbol{\Omega}_{k}(z,0)$ can be defined recursively as
\begin{equation}
  \begin{split}
    \frac{\der\boldsymbol{\Omega}_{k}(z,0)}{\der z}\hspace{-2pt}&=\hspace{-2pt}\begin{cases}
      \mathbf{A}(z),& k=1,\\
      \sum_{j=1}^{k-1}\frac{B_{j}}{j!}\mathbf{S}^{(j)}_{n}(z,0),& k>1,
    \end{cases}\\
    \mathbf{S}^{(j)}_{n}(z,0)\hspace{-2pt}&=\hspace{-2pt}\begin{cases}
      \left[\boldsymbol{\Omega}_{n}(z,0), \mathbf{A}(z)\right],& j=1,\\
      \sum_{p=1}^{n-j}\hspace{-2pt}\left[\boldsymbol{\Omega}_{p}(z,0), \mathbf{S}^{(j-1)}_{n-p}\right],\hspace{-8pt}& 1\hspace{-2pt}<\hspace{-2pt}j\hspace{-2pt}<\hspace{-2pt}n\hspace{-2pt}-\hspace{-2pt}1,\\
      \ad^{(n-1)}_{\boldsymbol{\Omega}_{1}}\left(\mathbf{A}(z)\right),& j=n-1,\\
    \end{cases}
  \end{split}
\end{equation}
where $B_{j}$ are the Bernoulli numbers with $B_{1}=\frac{1}{2}$, ${[\mathbf{A}, \mathbf{B}]=\mathbf{A}\mathbf{B}- \mathbf{B}\mathbf{A}}$ is the commutator operator, and the matrix $\mathbf{A}(z)$ is introduced for conciseness and is given by
\begin{equation}
  \mathbf{A}(z)=\boldsymbol{\alpha}_{s}+\mathbf{K}_{s}+\diag\left[\Gr\AeffMat\vec{P}_{p}(z)\right].
\end{equation}

Finally, the function $\ad^{(n)}_{\boldsymbol{\Omega}}(\cdot)$ is a linear operator defined iteratively by
\begin{equation}
  \ad^{(n)}_{\boldsymbol{\Omega}_{1}}\left(\mathbf{A}(z)\right)=\begin{cases}
    \mathbf{A}(z),& n=0,\\
    \left[\boldsymbol{\Omega}_{1},\ad^{(n-1)}_{\boldsymbol{\Omega}_{1}}\left(\mathbf{A}(z)\right)\right],& n>0.\\
  \end{cases}
\end{equation}

By truncating the series at a given term $K$, a practical approximate solution can be obtained. The approximation error depends on the truncation order and on the degree to which the loss and coupling matrices commute with the Raman gain term. For the first term, a closed-form solution can be derived and is given in Eq.~\eqref{eq:pcf1}.

To evaluate the approximation error, a suitable quantitative metric is required. A simple norm of the power-error vector at $z_{\mu}$ does not provide a general error bound, since it depends on the power distribution at $z_{\nu}$. Therefore, the error is defined as the norm of the logarithmic-scale difference between the solution truncated at order $K$ and the exact solution as
\begin{equation}
  \begin{split}
    \epsilon_{K}\hspace{-3pt}&=\hspace{-2pt}\left\lVert\ln\hspace{-2pt}\left\{\hspace{-2pt}\exp\hspace{-2pt}\left[\sum_{k=1}^{\infty}\boldsymbol{\Omega}_{k}(z,0)\right]\right\}\hspace{-2pt}-\hspace{-2pt}\ln\left\{\hspace{-2pt}\exp\hspace{-2pt}\left[\sum_{k=1}^{K}\boldsymbol{\Omega}_{k}(z,0)\right]\right\}\right\rVert\\
    &=\hspace{-2pt}\left\lVert\sum_{k=K+1}^{\infty}\boldsymbol{\Omega}_{k}(z,0)\right\rVert\leq \sum_{k=K+1}^{\infty}\lVert\boldsymbol{\Omega}_{k}(z,0)\rVert,
  \end{split}
\end{equation}
where $\lVert\cdot\rVert$ denotes an arbitrary sub-multiplicative matrix norm.

\begin{figure}[!t]
    \centering
    \includegraphics{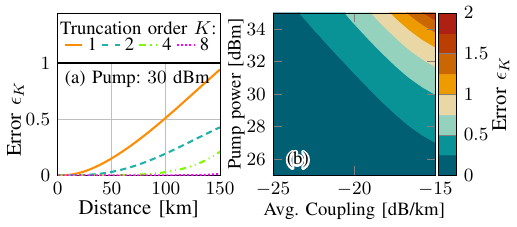}
    \caption{(a) Approximation error versus distance, where line styles indicate distinct truncation orders. (b) Error of the first-order approximation ($K=1$) versus pump power and average modal coupling. Curves correspond to a 3-mode-group \gls*{gi-mmf} with a co-propagating Raman pump allocated to the fundamental mode-group.}
    \label{fig:MagnusErr}
\end{figure}

Figure~\ref{fig:MagnusErr} shows how the approximation error varies with different parameters. The error curves are compared against the Magnus series truncated at $K=12$. The setup parameters are listed in Tab.~\ref{tab:parameters} and Fig.~\ref{fig:MmfParam}. For Fig.~\ref{fig:MagnusErr}(b), the coupling matrix in Fig.~\ref{fig:MmfParam}(c) is linearly scaled to achieve the desired mode-averaged crosstalk values.

Figure~\ref{fig:MagnusErr}(a) shows that the first-order error increases monotonically with distance, with different growth rates depending on the truncation order. Figure~\ref{fig:MagnusErr}(b) shows that the error increases significantly as either the pump power (and consequently the Raman gain) or the crosstalk becomes larger.

\subsection{Multi-section approximation}
\label{app:b1}

The norm of each term in the Magnus series is coarsely bounded by~\cite[Section 2.7.2]{blanes2009magnus},~\cite{moan2002backward}
\begin{equation}
  \lVert\boldsymbol{\Omega}_{k}(z,0)\rVert\leq \frac{\pi}{\xi^{k}}\left[\int_{0}^{z}\lVert A(z')\rVert_{2}\der z'\right]^{k}
\end{equation}
with $\xi\approx 1.08686870$.

By partitioning the fiber span into small sections, $\mathbf{A}(z)$ can be approximated as constant along the $\nu^{\Th}$ slice, such that ${\int_{z_{\nu}}^{z_{\mu}}\lVert A(z')\rVert_{2}dz'\approx \lVert A(z_{\nu})\rVert_{2}\Delta z}$ with ${\Delta z=z_{\mu}-z_{\nu}}$ being the section length. Under this approximation, the Magnus series terms for each section are bounded by
\begin{equation}
  \lVert\boldsymbol{\Omega}_{k}(z_{\mu},z_{\nu})\rVert\leq \frac{\pi}{\xi^{k}}\left[\int_{z_{\nu}}^{z_{\mu}}\lVert A(z')\rVert_{2}\der z'\right]^{k}\hspace{-8pt}\tilde{\equiv}\pi\frac{\left(\lVert A(z_{\nu})\rVert_{2}\Delta z\right)^{k}}{\xi^{k}},
\end{equation}
which bounds the error in each section by
\begin{equation}
  \epsilon_{K,\nu}\leq\pi\sum_{k=K+1}^{\infty}\frac{\left[\lVert A(z_{\nu})\rVert_{2}\Delta z\right]^{k}}{\xi^{k}}\equiv \pi\frac{\left(\lVert A(z_{\nu})\rVert_{2}\Delta z/\xi\right)^{K+1}}{1-\lVert A(z_{\nu})\rVert_{2}\Delta z/\xi}.
\end{equation}

\begin{figure}[!t]
    \centering
    \includegraphics{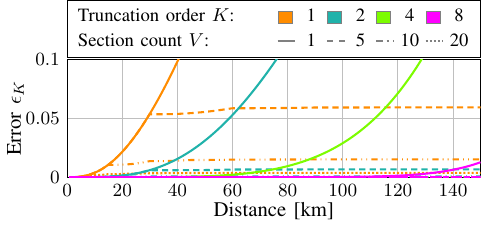}
    \caption{Approximation error versus distance for a 3-mode-group \gls*{gi-mmf} with a 30~dBm co-propagating pump allocated to the fundamental mode-group. Line colors indicate the truncation order, and line styles the number of sections.}
    \label{fig:MagnusSection}
\end{figure}

The accumulated error is not trivial to estimate, as the power at each section differs between the exact and approximate solutions. Nevertheless, in the small-error regime, which is the desirable scenario, the norm of the power evolution matrix is much greater than the error. Consequently, a small error in the evolution matrix for a given section does not significantly affect the error estimation in the following section, provided that the error decreases more than linearly with the section count, as shown subsequently. Under this assumption, the error accumulates additively across sections
\begin{equation}
  \epsilon_{K}\approx\sum_{\nu=1}^{V}\epsilon_{K,\nu}\equiv \sum_{\nu=1}^{V}\pi\frac{\left(\lVert A(z_{\nu})\rVert_{2}\Delta z/\xi\right)^{K+1}}{1-\lVert A(z_{\nu})\rVert_{2}\Delta z/\xi},
\end{equation}
where, for proper convergence, the section size should satisfy ${\Delta z_{\nu} < \lVert A(z_{\nu})\rVert_{2}\xi}$. The error scales roughly as $V^{-K}$. Since the error decay exponentially with the truncation order, one can choose to increase the number of Magnus series terms rather than increasing the number of sections. However, the Magnus terms become increasingly intricate at higher orders, as they involve nested adjoint commutators. By considering only the first-order Magnus solution, a closed-form expression involving matrix products can be obtained, but at the cost of requiring more sections to reduce the error. This solution is provided in Eq.~\eqref{eq:cfsplit}, and Fig.~\ref{fig:MagnusSection} illustrates how the error scales with different truncation orders and section counts.

\section{Zero-ISRS-induced frequency}
\label{app:c}

By multiplying all sides of~\eqref{eq:ptotal} by the inverse of the loss and crosstalk term and using the approximation in~\eqref{eq:cfisrs}, while assuming commutativity due to the small \gls*{srs} gain/loss, we obtain
\begin{equation}
  \vec{P}_{\Total}(0)=\int_{\fmin}^{\fmax}\exp\left[\left(\boldsymbol{\alpha}\hspace{-2pt}-\hspace{-2pt}\mathbf{K}\hspace{-1pt}\right)\hspace{-2pt}z+\boldsymbol{\tilde{\Omega}}_{1}(f,z)\right]\vec{S}(f,0).
  \label{eq:c1}
\end{equation}

The exponential term in Eq.~\eqref{eq:c1} is a diagonal matrix, so the equality can be evaluated element-wise as
\begin{equation}
  P_{\Total,n}(0)=\int_{\fmin}^{\fmax}\exp\left[\Cr(\Frm{n}-f)P_{\eff,n}(z)\right]S_{n}(f,0),
\end{equation}
where $P_{\eff,n}(z)$ is defined in Eq.~\eqref{eq:peff}. After some algebraic manipulation, an expression for $\Frm{n}$ is obtained as given in Eq.~\eqref{eq:fr}.

\bibliographystyle{IEEEtran}
\bibliography{references}

\end{document}